\documentclass[aps,pra,twocolumn,amsmath,amssymb,superscriptaddress,showpacs]{revtex4-1}

\usepackage{amsfonts}
\usepackage{amsmath}
\usepackage{amssymb}
\usepackage{amsthm}
\usepackage{fontenc}
\usepackage{graphicx}
\usepackage{xcolor}
\usepackage{textcomp}
\usepackage{epstopdf}
\usepackage{braket}
\usepackage{mathtools}
\usepackage{amsmath}
\usepackage{dcolumn}
\usepackage{multirow}
\usepackage{units}

\begin{document}
\newcommand{\abs}[1]{\lvert#1\rvert}

\title{Topological phases of graphene-Kagome systems}%possible title

\author{A. B. Felix de Souza}
\affiliation{Instituto de F\'isica, Universidade Federal Fluminense, Niter\'oi, Av. Litor\^{a}nea sn 24210-340, RJ-Brazil}
\author{L. Spreafico}
\affiliation{Instituto Polit\'ecnico, Universidade do Estado do Rio de Janeiro, Nova Friburgo, RJ, Brazil}

\author{D. Faria}
\affiliation{Instituto Polit\'ecnico, Universidade do Estado do Rio de Janeiro, Nova Friburgo, RJ, Brazil}

\author{A. Latgé}
\affiliation{Instituto de F\'isica, Universidade Federal Fluminense, Niter\'oi, Av. Litor\^{a}nea sn 24210-340, RJ-Brazil}

\date{\today}

\begin{abstract}

The growing skill in the synthesis processes of new materials has intensified the interest in exploring the properties of systems modeled by more complex lattices. Two-dimensional super-honeycomb lattices, %including the graphene-Kagome systems,
have been investigated in metallic organic frameworks. 
They turned out as a significant
route to the emergence of localized electronic responses, manifested as flat bands in their
structure with topological isolating behavior.
 A natural inquiry is a complete analysis of their topological phases in the presence of electronic correlation effects. Here  we analyse of the electron-electron correlation effects via Hubbard mean-field approximation on the topological phases of 2D and quasi-1D graphene-Kagome lattices. The 2D spin conductivity phase’s diagrams describe metallic, trivial and topological insulating behaviors, considering different energy coupling 
and electronic occupations. Our results pave the way to smart-engineered nanostructured devices with relevant applications in spintronics and transport responses.

\end{abstract}

\maketitle

%---------------------------------------------
\section{Introduction}
%---------------------------------------------

Kagome lattices have been largely explored within the context of covalent frameworks \cite{Chachan2016}. Metallic-Organic Frameworks (MOFs) appeared as a new generation of ultrahigh porosity systems with a large and high internal surface area \cite{Zhou2012}, allowing fine-tuning of the structures and their properties through, for instance, the adsorption of molecules by the pores of the network\cite{Talin2014,Campbell2015}.

Different studies reveal interesting electronic properties in 2D metal-organic materials, such as superconductivity\cite{Takenaka2021}, Dirac cones \cite{Wu2014} and half-metallicity\cite{Hu2014}. In particular, theoretical proposals of 2D MOFs of Archimedean structure present gaps induced by spin-orbit effects and non-trivial topological phases\cite{felipe45}. Actually, topological properties of a variety of 2D metal–organic Kagome-honeycomb lattice were reported\cite{Sun2018,Wang2013}, addressing the possibility of tuning spin-orbit coupling intensities and the transport properties by a proper selection of vertices and bonds of the hybrid systems. In that sense, Rashba and intrinsic spin-orbit couplings provide additional tools for designing 2D-Topological Insulator (TI) band structures. It is important to mention that complex Kagome lattices have also been reported as basic photonic crystals leading to important higher-order topological phases \cite{Khanikaev2020,Kirsch2021}. 

In general, the synthesis of these materials occurs through chemical polymerization protocols of organic monomers\cite{Cho2015,Xu2015,Dienstmaier2011, Zhang2013,Feldblyum2015}, whose size and shape strongly depend on the physicochemical parameters of the solution (solvent, pH, concentration, temperature, pressure, etc.)\cite{Stock2012}. Bottom-up and top-down methods\cite{Tan2019} have been used to sinthetize low-dimensional crystalline MOFs such as nanoparticles (0D), nanoribbons and nanotubes (1D), 2D-nanosheets and -nanoflakes. The spatial geometry of the metal-organic framework is highly dependent on the geometry details of the binding organic molecules. Therefore, a judicious choice of molecules and metal ions can lead to the synthesis of two-dimensional crystalline structures with desired properties\cite{Iqbal2016}. Non-trivial 1D topological phases were addressed in experimental realization of precise all-graphene nanoribbon hetrostructures, opening  new promising routes for band engineering based on controlling their electronic  topology \cite{Rizzo2018}. Some of the applications for such materials involve energy storage devices (batteries and supercapacitors) \cite{Sheberla2017} and sensors \cite{Xang2018}. 

A simplified theoretical description of such MOF systems may be addressed via tight-binding models on the Kagome lattice. In general, they predict a dispersionless band in the energy spectrum, not present in the hexagonal graphene lattices that has been extensively explored \cite{Montambaux2020,Essafi2017,Rhim2019}.
When spin-orbit coupling are also taken into account these flat bands may exhibit interesting topological features\cite{Liu2009,Leykam2018,Boleans2019}, due to the occurrence of degeneracy breaking. Magnetic ordering features in Kagome systems with transition metals has inspired theoretical work on standard Kagome lattices including strong electronic correlations via Hartree-Fock approach \cite{Kim2020}. Observation of topological flat bands in frustrated kagome metal CoSn were reported using angle-resolved photoemission spectroscopy and band structure calculations\cite{Arpes2020}.

Here we explore the conjugated effects of electron-electron interaction and spin-orbit coupling in the topological properties of hexagonal graphene-Kagome lattices. We first analyze the topological emergence on the isolated couplings, i.e., spin-orbit with null e-e interaction and also electronic correlation without spin-orbit coupling. We then consider both interactions. One of the theoretical findings is that although topological transitions happen  for high spin-orbit coupling, the incorporation of electronic correlation may reduce the transition SOC value. We also observe that electronic correlation alone promotes changes in the band topological characterization  that depends also on the filling band factor. Further, we explore  the topological nature of the edge states by calculating the probability density of the electronic wave functions.
The edge states are shown to counterpropagates with reverse spin in the opposite edges of the graphene-kagome nanoribbon.

%---------------------------------------------
\section{Model}
%--------------------------------------------. 

The graphene-Kagome systems are modeled by a single-orbital Hubbard like tight-binding Hamiltonian written as 

\begin{eqnarray}\label{eq:1}
    H&=&\sum_{ i,\alpha} \varepsilon_{i}c^{\dagger}_{i,\alpha}c_{i,\alpha} +\sum_{ \langle i,j \rangle,\alpha} t_{i,j}c^{\dagger}_{i,\alpha}c_{j,\alpha}+
%\nonumber\\ 
   \sum_{ i,\alpha} U_{i} \langle n_{i,\bar{\alpha}} \rangle n_{i,\alpha} + \nonumber\\
   &+& i \lambda_{SO}\sum_{ \langle\langle i,j\rangle\rangle, \alpha, \beta} \nu_{i,j}c^{\dagger}_{i,\alpha}\sigma^z_{\alpha,\beta}c_{j,\beta}+  h.c.,
\end{eqnarray}
where the first two terms are the usual on-site, and first-neighbor hopping terms, respectively, with the site energy given by  $\varepsilon_i$ and the hopping parameter by $t_{i,j}$. $\alpha$ and $\beta$ denote the spin projections along the $z$-direction, perpendicular to the graphene-Kagome lattice. The second term is the mean-field Hubbard term, where $n_{i,\alpha}$ is the mean-occupation number of site $i$ for spin projection $\alpha$, up to the Fermi energy $(E_F)$. 
 Here the mean-occupation number is obtained self-consistently via $n_{i,\alpha}= \int_{-\infty}^{E_F} \rho_i^{\alpha} (E)dE$, with $\rho_i^{\alpha}$ being the local density of state. The last term is a Kane-Mele term\cite{Kane2005}, where $\nu_{i,j}=\pm 1$, depending on the orientation of the two second neighbor bonds \cite{Kane_2005}, $\sigma^z_{\alpha,\beta}$ is the Pauli matrix element, and $\lambda_{SO}$ is the intensity of the intrinsic spin-orbit coupling (ISOC). All the energies are given in terms of the hopping parameter $t_{i,j}=t$.

\begin{figure}
\centering
 \scalebox{0.265}{
\includegraphics{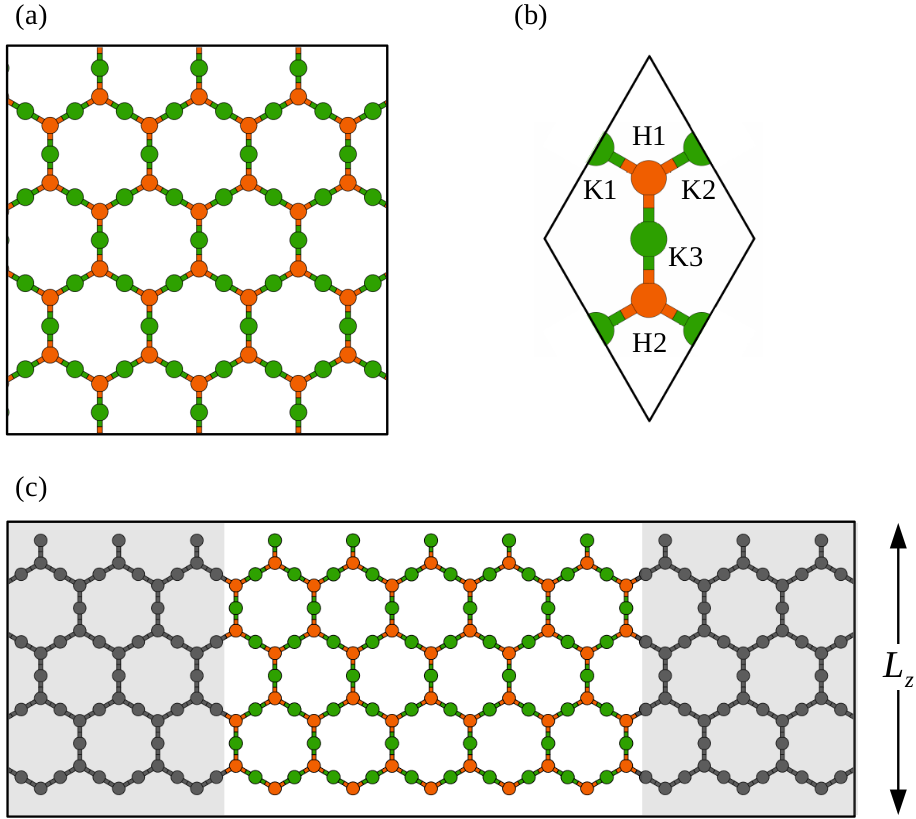}}
    \caption{Schematic view of (a) a 2D graphene-Kagome lattice composed of two different sites (orange and green) and (b) the corresponding unit cell. (c) A quasi 1D asymmetric graphene-Kagome nanoribbon with zigzag edges and width $L_z$. }
\label{figlattice}
\end{figure}

A 2D graphene-Kagome lattice is schematically displayed in Fig.\ref{figlattice} (a) with its corresponding unit cell in Fig.\ref{figlattice} (b). We also consider graphene-Kagome nanoribbons with zigzag edges displayed along the $x$-direction (ZGKNRs), with symmetric and asymmetric edges configurations. We call symmetric edged nanoribbons those with equal bottom and top edge configurations, differently from the asymmetric  geometry shown in Fig. \ref{figlattice} (c) where the central conductor is connected to two perfectly matched leads composed of the same nanoribbon nature. The symmetric (asymmetric)  nanoribbon width is given by $L_Z=10p-1$ ($L_Z=10p$), being $p$ the number of transversal hexagon units in the ribbon unit cell.  

In the case of the 2D system, a $N \times N$ eigenvalue problem defined by a matricial Hamiltonian is solved self-consistently,  using a standard definition of the discretized local density of states, with N being the number of atoms in the unit cell. Alternatively, for the graphene-Kagome nanoribbons the electronic density of states is derived via real-space renormalization techniques and standard recursive methods within the Green function formalism\cite{Rosales2008,carrillo2016strained,torres2018}. The local density of states is given as a function of energy and the local mean occupation number for the zigzag nanoribbon is also obtained selfconsistently\cite{Leon2019}.

\begin{figure}
   \centering
   \scalebox{0.23}{\includegraphics{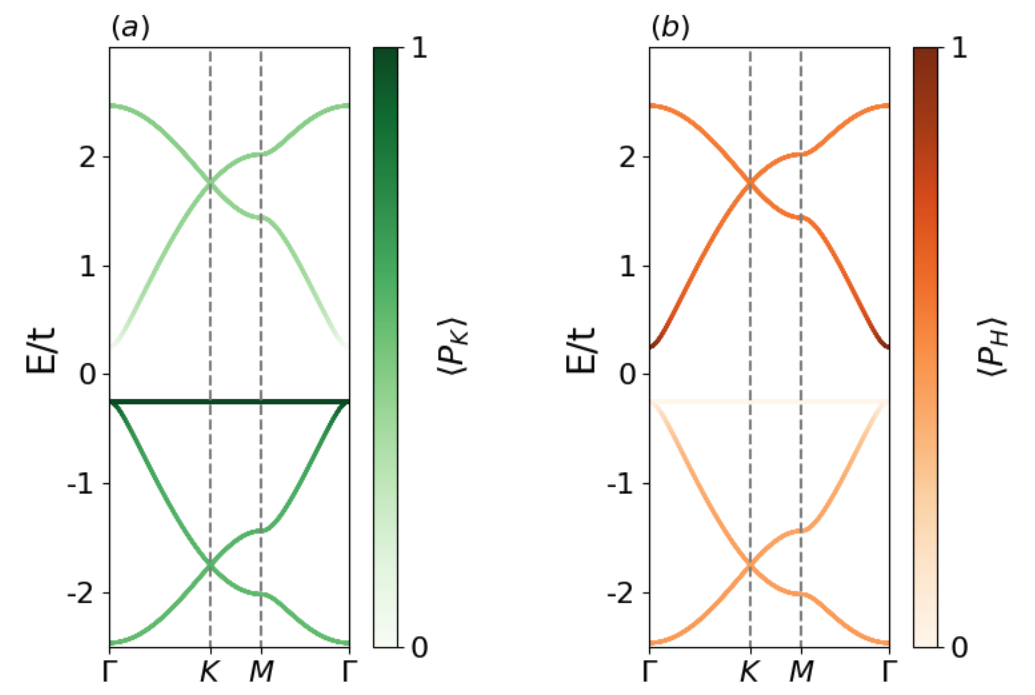}}
   \caption{Electronic band structure of a 2D graphene-Kagome lattice projected on both K (left) and H (right) sublattices along the symmetric lines of the Brillouin zine, with $\epsilon=\pm0.25t$ and null e-e and spin-orbit interactions.}
   \label{band0}
\end{figure}

Results of the electronic band structure of a 2D graphene-Kagome lattice projected on K and H sublattices  are  shown in Fig.\ref{band0} (a) and (b), respectively, for an uncorrelated 2D lattice and considering a null ISOC, i.e., U=0 and $\lambda_{SO}$=0. The electronic band structure gather features from both sublattices, i.e., a flat band and two pairs of graphene-like bands. The band gap at the $\Gamma$ point is given by $|\varepsilon_{K} - \varepsilon_{H}|$, where $\varepsilon_{K(H)}$ are on-site energies of K(H) sublattice. The comparison between the two sublattice projected band structures reveals a flat band built essentially by K-lattice states.

A topological characterization of the electronic states is also performed by calculating the spin Chern number, by simply integrating the Berry curvature for each energy band in the system's Brillouin zone \cite{Sheng2006}. The Berry curvature may be expressed as 
\begin{equation}\label{Berry_Curv}
    B_{n,\alpha} = i \sum_{n' \neq n} \frac{\langle n, \alpha | [ \nabla \hat{H}] | n', \alpha \rangle \times \langle n', \alpha | [ \nabla \hat{H}] | n, \alpha \rangle}{(E_{n, \alpha} - E_{n', \alpha})^{2}}
\end{equation}
with $E_{n\alpha}$ being the energy eigenvalue of the $n^{th}$ band. The spin Chern number is obtained directly from the Berry curvature $ C_{n, \alpha} = \frac{1}{2\pi} \int B_{n, \alpha} d^{2}k$, with $n$ indicating the different valence bands. The spin-dependent Hall conductivity is directly obtained via
\begin{equation}
    \sigma_{\alpha} = \frac{e^{2}}{h} \sum_{n=1}^{N} C_{n, \alpha}\,\,.
\end{equation}

%---------------------------------------------
\section{Results} 
\begin{figure}
   \centering
   \scalebox{0.35}{
   \includegraphics{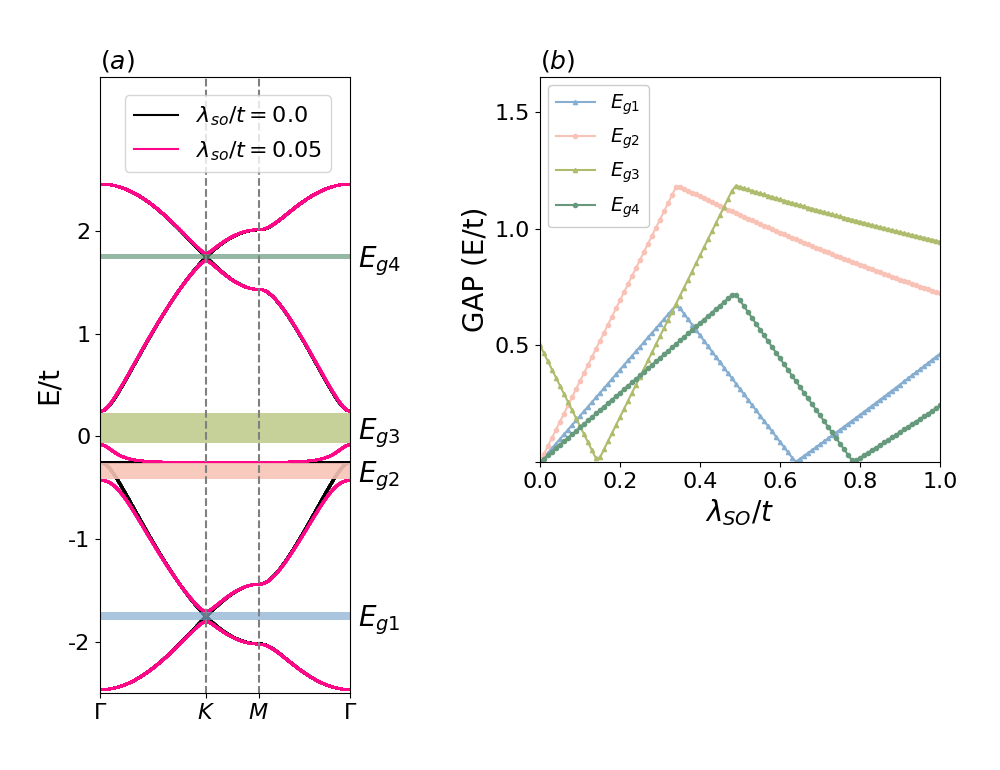}}
    \caption{(a) Band structure of 2D graphene-Kagome for $\lambda_{SO}/t = 0$ (black curves) and $\lambda_{SO}/t = 0.05$ (pink curves). (b) Energy gap $E_{gi}$ evolution as a function of intrinsic spin-orbit coupling parameter.}
    \label{fig:band2DSO}
\end{figure}

%---------------------------------------------
We first present the electronic band structure of an uncorrelated 2D graphene-Kagome lattice with a spin-orbit coupling $\lambda_{SO}/t$=0.05, as shown in  Fig.\ref{fig:band2DSO}(a) and null electron-electron correlation. For comparison we have also included the case of zero SO coupling. The SO coupling is responsible for important gap opening at K and $\Gamma$ points, named here as $E_{g1}$, $E_{g2}$ and $E_{g4}$. The gaps are marked by colored shadow regions. Near the $\Gamma$ point the flat band acquires an energy dispersion, modifying the gap size at E = 0 ($E_{g3}$). The gap evolution as a function of the SO coupling  for the different gaps are shown in Fig.\ref{fig:band2DSO}(b).

 The distinct behavior of the $E_{g1}$ to $E_{g4}$ gap curves put in evidence that intrinsic spin-orbit coupling breaks the electron-hole symmetry. In particular, the dependence with increasing $\lambda_{SO}$ values for both  $E_{g1}$ and $E_{g4}$ gap shows oscillatory-like features. Differently, the finite central gap  ($E_{g3}$), presented at $\lambda_{SO}=0$, closes for small values  of the spin-orbit interaction but evolves increasing it again, similarly to the features exhibited by the new SO-dependent gap $E_{g2}$ emerged when the flat band allow energy dispersion. 

\begin{figure}
   \centering
   \scalebox{0.53}{
   \includegraphics{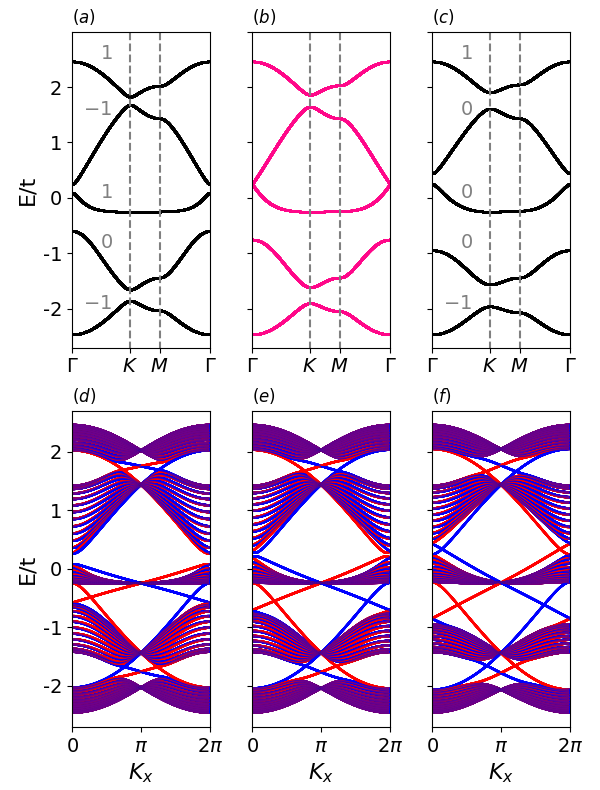}}
    \caption{Electronic band structures for 2D graphene-Kagome lattices considering $\epsilon=\pm0.25t$ and (a) $\lambda_{SO}/t =$ 0.100, (b) 0.144 and (c) 0.200. Spin-up Chern numbers for each band are depicted. Electronic energy bands for a 61-ZGKNR (d)-(f) for the same $\lambda_{SO}$ values as the 2D, and a filling factor equal to 3/5.}
    \label{fig:bands}
\end{figure}

In Fig.\ref{fig:bands} we display the electronic band structures for 2D graphene-Kagome lattices considering null electronic correlation and different  spin-orbit interaction, including the case at which the central gap is closed  ($\lambda_{SO}/t =$ 0.100, 0.144 and  0.200). The spin-up Chern numbers for each band are depicted, highlighting  the changes as a function of increasing $\lambda_{SO}$ values as the spin-orbit interaction is switched on. Also, the flat band is distorted at the $\Gamma$ point leading to a graphene-Kagome transition from a trivial to a topological insulator, that is inter-mediated by the situation of closed gap shown in figure \ref{fig:bands} (b)  for $\lambda_{SO}=0.14435$. 
Electronic energy bands for an asymmetric  60-ZGKNR are displayed in  figure \ref{fig:bands}(d)-(f) for the same $\lambda_{SO}$ values used in the example of the the 2D graphene-Kagome. The spin degeneracy is clearly lifted for finite $\lambda_{SO}$ values giving rise to surface states at the corresponding energy gap regions of the 2D counterpart system.  

\begin{figure}
   \centering
   \scalebox{0.55}{
   \includegraphics{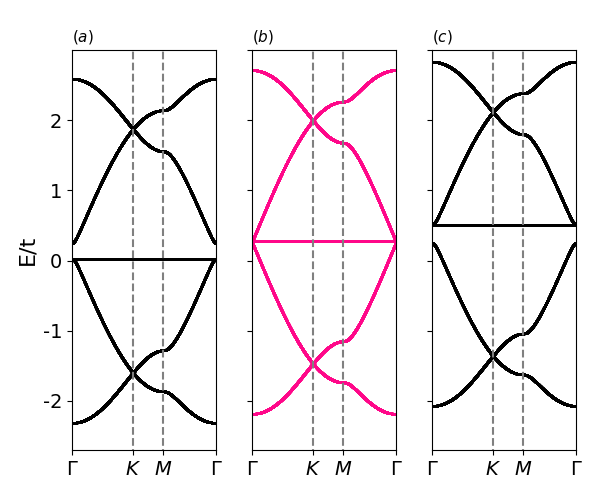}}
    \caption{Electronic band structures for 2D graphene-kagome lattice with $\lambda_{SO}/t = 0.0$ and (a) $U/t = 0.4$, (b) $U/t = 0.8$ and (c) $U/t = 1.2$. The electronic filling factor is 3/5 filling.}
    \label{fig:bandxx}
\end{figure}

Coming back to the 2D lattice we show in Fig. \ref{fig:bandxx} the effect on the electronic band structures by including electron-electron correlation in the Hamiltonian. All the results refer to the case of zero spin-orbit coupling  and considering an electronic filling factor $\zeta$=3/5 for different values of $U/t$. Other $\zeta$ values may also be explored, mainly those related to energy gap regions, as 1/5, 2/5 and 4/5.  No split of the spin bands are observed, imposing nonmagnetic solutions. As the e-e interaction increases the system suffers a semiconductor-metal transition. Interestingly, the gap changes from the top of the valence band to the bottom of the conduction band, guided by the position of the flat band. We have also analyzed the changes induced by the e-e correlations on the electronic properties of the zigzag ZGKNRs. The main feature found was the degeneracy lift of the flat band, highly pronounced  for increasing values of electronic correlation  and more notable in the case of asymmetric nanoribbons (not shown here). As in the 2D case, no spin degeneracy was manifested.

\begin{figure}
   \centering
   \scalebox{0.30}{
   \includegraphics{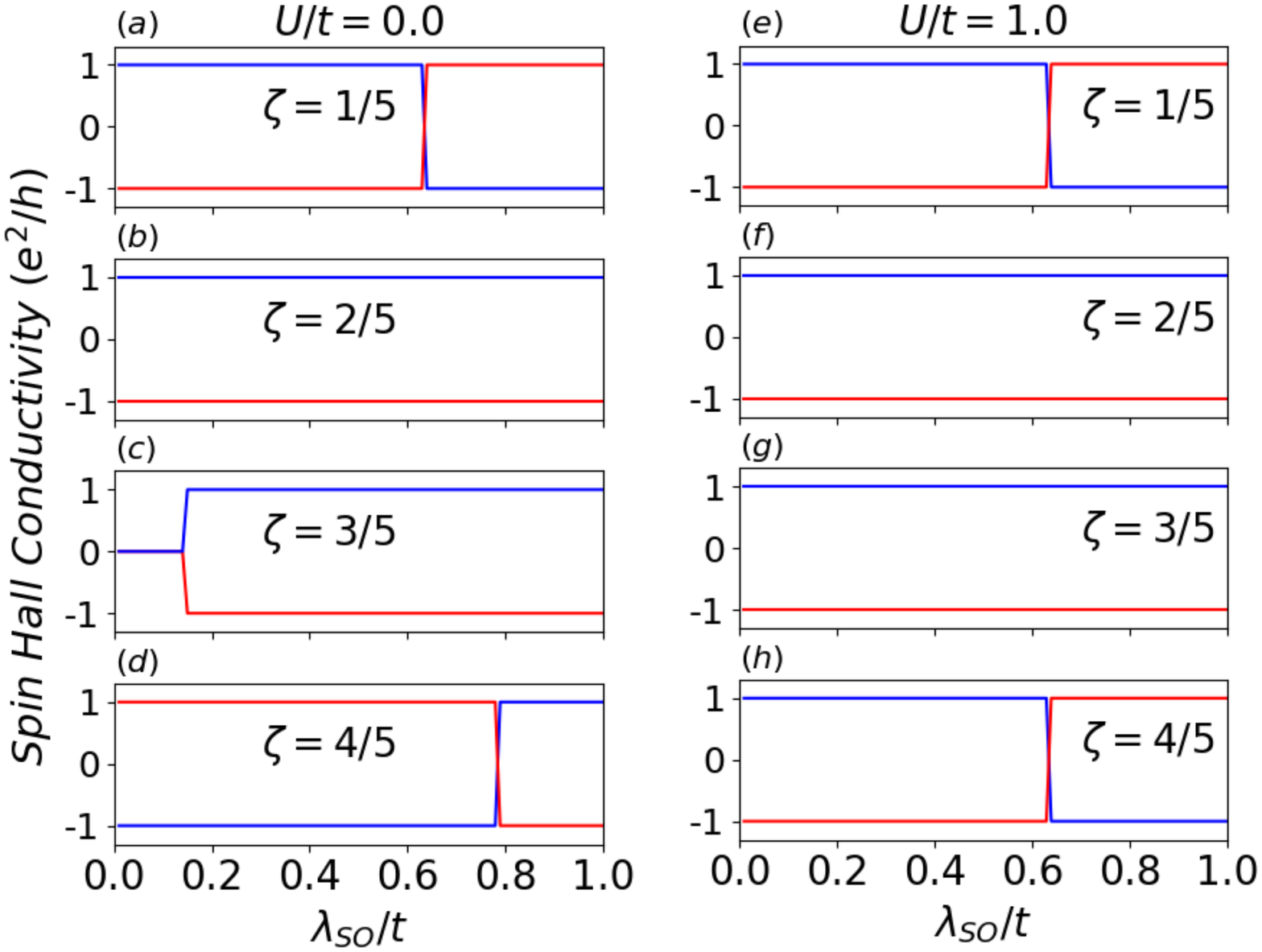}}
    \caption{Panels of spin polarized conductivity  as a function of the energy of the spin-orbit coupling ($\lambda_{SO}$) for different $\zeta$ values, considering $U=$0.0 and 1.0t.}
    \label{spincond}
\end{figure}

\begin{figure}
   \centering
   \scalebox{0.345}{
   \includegraphics{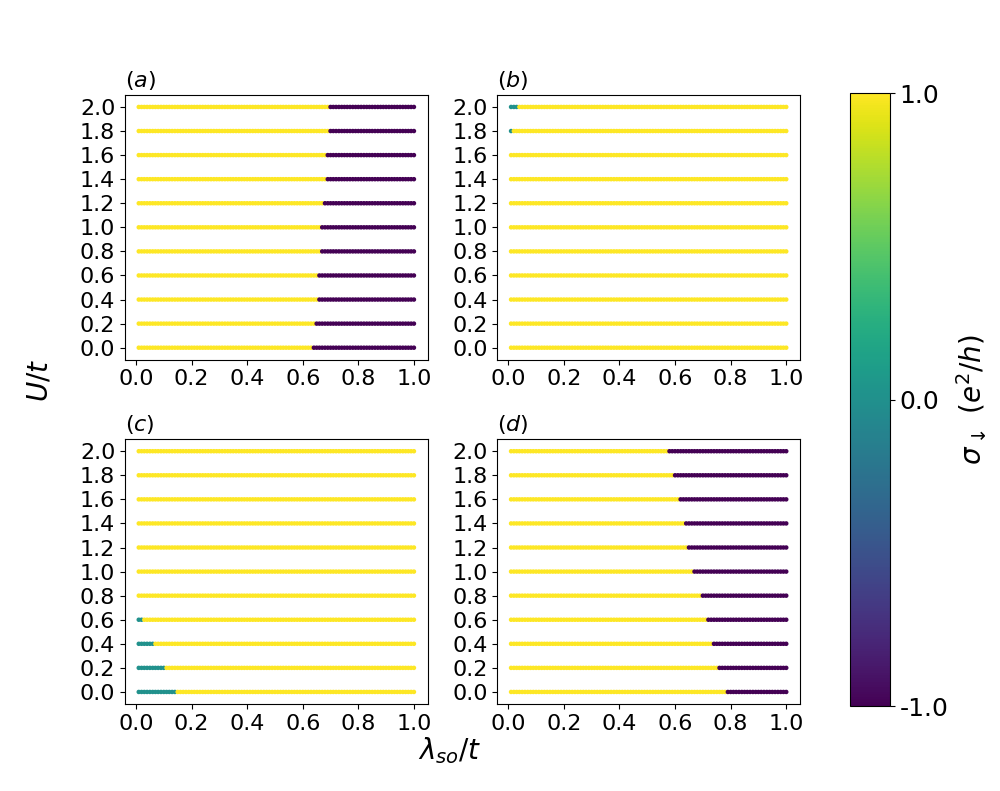}}
    \caption{Diagrams of spin polarized conductivity maps as a function of the spin-orbit and e-e interaction energies, for (a) 1/5, (b) 2/5, (c) 3/5  and (d) 4/5 filling factors.}
    \label{mosaico}
\end{figure}

\begin{figure}
   \centering
   \scalebox{0.375}{
   \includegraphics{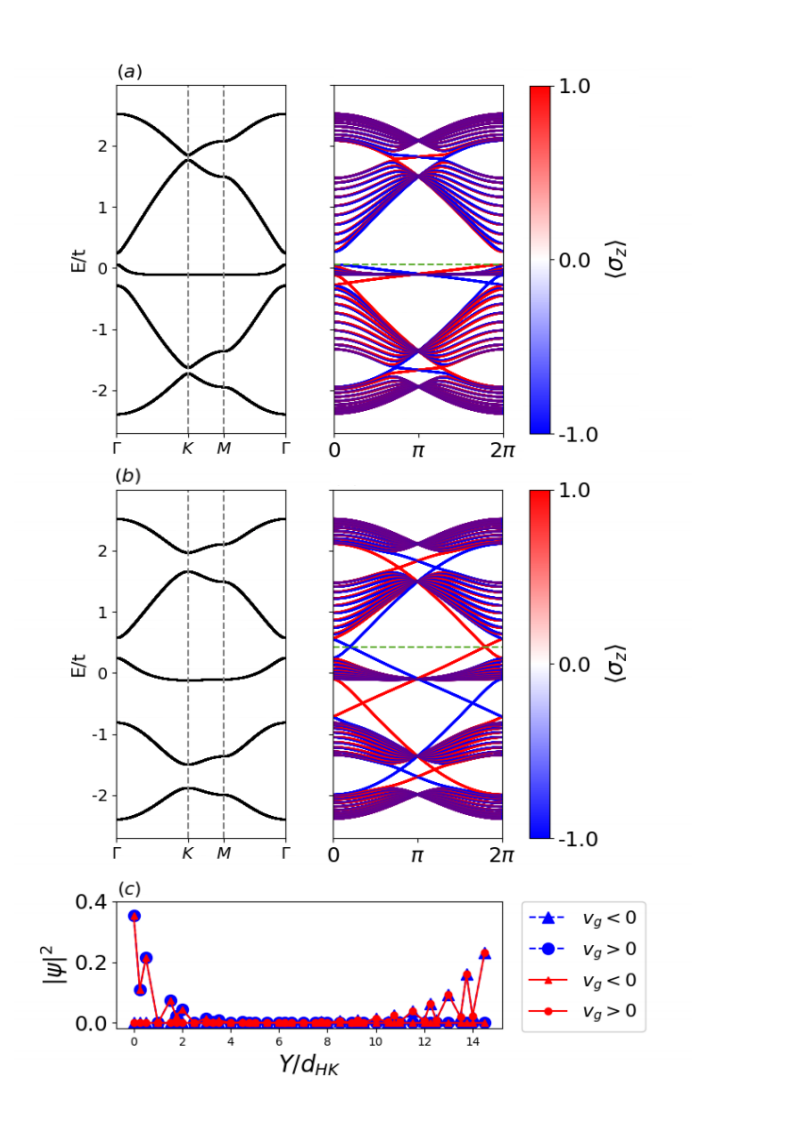}}
    \caption{Band structure of an asymmetric 50-ZGKNR with $U/t = 0.2$ and (a) $\lambda_{so}/t = 0.05$ and (b) $\lambda_{so}/t = 0.2$, for the case of $\zeta=$ 3/5. The Fermi level is marked with green dashed line. Red and blue curves denote up and down spin components, respectively. Band structures of the corresponding 2D Kagome lattices are displayed at right. (c) Probability density of metallic states for $\lambda_{so}/t = 0.2$ at $E/t=0.5$, along the transversal direction, given in units of H-K NN distance, $d_{HK}$ }
    \label{fig:walkstates}
\end{figure}

The  dependence of the spin-polarized conductivity on the spin-orbit coupling, calculated via Eq.(3), is presented in Fig.\ref{spincond} for null e-e correlation (left panel) and for $U/t=1.0$ (right panel), considering different values of $\zeta$. For null e-e correlation, three cases emerge depending on the filling factor: (i) normal semiconductor to topological insulator transition ($\zeta$=3/5), and no transition from trivial to topological (and vice-versa) with (i) spin-polarized conductivity changes or with (ii) fixed spin polarization for all the SO coupling range studied.  However, for the particular case of U=1.0t, only situations (ii) and (iii) are manifested, i.e., no topological transition is achieved as the spin-orbit is increased. Actually, while for $\zeta$ equal to 1/5, 2/5 and 4/5 there are not great differences between the spin-polarized conductivity results with and without the electronic correlations, the doped case corresponding to a filling equal to 3/5 marks a phase transition at a particular SO coupling value. It should be interesting, therefore, to investigate  the dependence of the  spin-polarization conductivity on both correlation and coupling energies.
 
The spin Chern number changes as the electron-electron interaction and the spin-orbit coupling are considered are illustrated on the colored diagrams displayed in Fig.\ref{mosaico} via the conductivity results. We used Eq. (3), taking into account that the number of edge metallic states corresponds to the sum of the spin Chern number of the occupied bands.  Depending on the filling factor, the transition from topological insulators to trivial insulator, and vice-versa, occurs (or not) for particular combinations of spin-orbit coupling and e-e interaction values. This is revealed by the color change in the diagrams. In the case of $\zeta=$3/5, such transition is promoted for the lowest coupling energy values; $\lambda_{SO}<0.2t$ and $U<0.7t$, green to yellow changes in the diagram. This is particularly considered an interesting case because of the relative low value of the spin-orbit coupling necessary to promote the topological insulator transition.

Both interaction effects (e-e correlation and spin-orbit coupling) are also explored on the ZGKNR's. The electronic structures of asymmetric 60-ZGKNR with $\zeta=$ 3/5, $U/t=0.2$, and $\lambda_{so}/t=0.05$ and $0.20t$ are shown in Fig. \ref{fig:walkstates}(a) and (b), respectively. The Fermi level for both cases are marked with green dashed lines. The band structures of the corresponding 2D lattices with the same energy parameters (U and $\lambda_{SO}$) are shown at right. When the spin-orbit coupling is small [see case (a) in Fig. \ref{fig:walkstates}] the Fermi level clearly marks the presence of a normal gap, typical of a trivial insulator. Contrary, for $\lambda_{so}/t=0.2$, two groups of crossing bands emerge, each for one k-direction,  separating the opposite spin states within the topological gaps.  The presence of such edge states reveals the topological insulator nature of the graphene-Kagome nanoribbon systems. To highlight this feature,  we display in Fig.\ref{fig:walkstates}(c) the spatial distribution of the wave function associated to an energy state close to the Fermi level (0.5t) corresponding to the Hamiltonian parameter used in Fig.\ref{fig:walkstates}(b). The position from one edge to the other was given in terms of the distance between two sites H and K in the unit cell, $d_{HK}$. The probability density associated with the different combinations of momentum and spin directions, denoted by distinct symbols, reveals the edge localization of the states typical of a topological insulator phase. 
It is important to mention that the
emergence of topological phases does not depend on the spatial symmetry but entirely on ISO effect. Symmetric nanoribbons also exhibit edge localizations but differently, the spin degeneracy lift is a subtle effect.

%---------------------------------------------
\section{Conclusion}
%---------------------------------------------

A topological insulator characterization of metal-organic frameworks, described by 2D graphene-Kagome systems and quasi 1D graphene-Kagome nanoribbons, were addressed by using a Hubbard like tight-binding Hamiltonian with spin-orbit coupling and the invariant Chern number per spin. Hall conductance diagram maps were derived,  for different electronic occupation and combined e-e correlation and SO coupling intensities. Trivial to non-trivial insulator phase transitions were obtained for 2/5 and  3/5 filling factors. For the case of the graphene-Kagome nanoribbons the topological insulator phases were analyzed from the emergence of metallic states at the central bulk gap,  and the corresponding  edge spatial localization of such states.

As the effects of considering both interaction couplings on the electronic responses of metal-organic systems, and their consequent topological nature, are so far little explored, we believe that our findings are quite relevant for future applications, taking into account the wide variety of MOFs already existing, synthesized and theoretically proposed.  Actually we have shown that graphene-Kagome is a topological insulator and depending on the spin-orbit considered topological transitions may occurs. Moreover, such topological transitions are also modulated by the electron-electron interaction intensities. The study of such lattices as describing real materials is fundamental  for their applicability as possible spintronic devices and as a source of possible routes of exotic material states. 

%---------------------------------------------
\section*{Acknowledgments}
%---------------------------------------------

This work was financially supported by the Brazilian Agencies CAPES, CNPq, and FAPERJ under the grant E-26/202.567/2019, and the INCT de Nanomateriais de Carbono. We would like to thank H. Chacham and H. Hiroki for fruitful discussions, and also R. Amorim and M. Corradini for helping  on the code optimizations.

\bibliography{ref}

\begin{thebibliography}{38}%
\makeatletter
\providecommand \@ifxundefined [1]{%
 \@ifx{#1\undefined}
}%
\providecommand \@ifnum [1]{%
 \ifnum #1\expandafter \@firstoftwo
 \else \expandafter \@secondoftwo
 \fi
}%
\providecommand \@ifx [1]{%
 \ifx #1\expandafter \@firstoftwo
 \else \expandafter \@secondoftwo
 \fi
}%
\providecommand \natexlab [1]{#1}%
\providecommand \enquote  [1]{``#1''}%
\providecommand \bibnamefont  [1]{#1}%
\providecommand \bibfnamefont [1]{#1}%
\providecommand \citenamefont [1]{#1}%
\providecommand \href@noop [0]{\@secondoftwo}%
\providecommand \href [0]{\begingroup \@sanitize@url \@href}%
\providecommand \@href[1]{\@@startlink{#1}\@@href}%
\providecommand \@@href[1]{\endgroup#1\@@endlink}%
\providecommand \@sanitize@url [0]{\catcode `\\12\catcode `\$12\catcode
  `\&12\catcode `\#12\catcode `\^12\catcode `\_12\catcode `\%12\relax}%
\providecommand \@@startlink[1]{}%
\providecommand \@@endlink[0]{}%
\providecommand \url  [0]{\begingroup\@sanitize@url \@url }%
\providecommand \@url [1]{\endgroup\@href {#1}{\urlprefix }}%
\providecommand \urlprefix  [0]{URL }%
\providecommand \Eprint [0]{\href }%
\providecommand \doibase [0]{http://dx.doi.org/}%
\providecommand \selectlanguage [0]{\@gobble}%
\providecommand \bibinfo  [0]{\@secondoftwo}%
\providecommand \bibfield  [0]{\@secondoftwo}%
\providecommand \translation [1]{[#1]}%
\providecommand \BibitemOpen [0]{}%
\providecommand \bibitemStop [0]{}%
\providecommand \bibitemNoStop [0]{.\EOS\space}%
\providecommand \EOS [0]{\spacefactor3000\relax}%
\providecommand \BibitemShut  [1]{\csname bibitem#1\endcsname}%
\let\auto@bib@innerbib\@empty
%</preamble>
\bibitem [{\citenamefont {Silveira}\ \emph {et~al.}(2016)\citenamefont
  {Silveira}, \citenamefont {Alexandre},\ and\ \citenamefont
  {Chacham}}]{Chachan2016}%
  \BibitemOpen
  \bibfield  {author} {\bibinfo {author} {\bibfnamefont {O.~J.}\ \bibnamefont
  {Silveira}}, \bibinfo {author} {\bibfnamefont {S.~S.}\ \bibnamefont
  {Alexandre}}, \ and\ \bibinfo {author} {\bibfnamefont {H.}~\bibnamefont
  {Chacham}},\ }\href {\doibase 10.1021/acs.jpcc.6b05081} {\bibfield  {journal}
  {\bibinfo  {journal} {J. Phys. Chem. C}\ }\textbf {\bibinfo {volume} {120}},\
  \bibinfo {pages} {19796} (\bibinfo {year} {2016})}\BibitemShut {NoStop}%
\bibitem [{\citenamefont {Zhou}\ \emph {et~al.}(2012)\citenamefont {Zhou},
  \citenamefont {Long},\ and\ \citenamefont {Yaghi}}]{Zhou2012}%
  \BibitemOpen
  \bibfield  {author} {\bibinfo {author} {\bibfnamefont {H.-C.}\ \bibnamefont
  {Zhou}}, \bibinfo {author} {\bibfnamefont {J.~R.}\ \bibnamefont {Long}}, \
  and\ \bibinfo {author} {\bibfnamefont {O.~M.}\ \bibnamefont {Yaghi}},\ }\href
  {\doibase 10.1021/cr300014x} {\bibfield  {journal} {\bibinfo  {journal} {Chem
  Rev.}\ }\textbf {\bibinfo {volume} {112}},\ \bibinfo {pages} {673} (\bibinfo
  {year} {2012})}\BibitemShut {NoStop}%
\bibitem [{\citenamefont {Talin}\ \emph {et~al.}(2014)\citenamefont {Talin},
  \citenamefont {Centrone}, \citenamefont {Ford}, \citenamefont {Foster},
  \citenamefont {Stavila}, \citenamefont {Haney}, \citenamefont {Kinney},
  \citenamefont {Szalai}, \citenamefont {Gabaly}, \citenamefont {Yoon},
  \citenamefont {Léonard},\ and\ \citenamefont {Allendorf}}]{Talin2014}%
  \BibitemOpen
  \bibfield  {author} {\bibinfo {author} {\bibfnamefont {A.~A.}\ \bibnamefont
  {Talin}}, \bibinfo {author} {\bibfnamefont {A.}~\bibnamefont {Centrone}},
  \bibinfo {author} {\bibfnamefont {A.~C.}\ \bibnamefont {Ford}}, \bibinfo
  {author} {\bibfnamefont {M.~E.}\ \bibnamefont {Foster}}, \bibinfo {author}
  {\bibfnamefont {V.}~\bibnamefont {Stavila}}, \bibinfo {author} {\bibfnamefont
  {P.}~\bibnamefont {Haney}}, \bibinfo {author} {\bibfnamefont {R.~A.}\
  \bibnamefont {Kinney}}, \bibinfo {author} {\bibfnamefont {V.}~\bibnamefont
  {Szalai}}, \bibinfo {author} {\bibfnamefont {F.~E.}\ \bibnamefont {Gabaly}},
  \bibinfo {author} {\bibfnamefont {H.~P.}\ \bibnamefont {Yoon}}, \bibinfo
  {author} {\bibfnamefont {F.}~\bibnamefont {Léonard}}, \ and\ \bibinfo
  {author} {\bibfnamefont {M.~D.}\ \bibnamefont {Allendorf}},\ }\href {\doibase
  10.1126/science.1246738} {\bibfield  {journal} {\bibinfo  {journal}
  {Science}\ }\textbf {\bibinfo {volume} {343}},\ \bibinfo {pages} {66}
  (\bibinfo {year} {2014})}\BibitemShut {NoStop}%
\bibitem [{\citenamefont {Campbell}\ \emph {et~al.}(2015)\citenamefont
  {Campbell}, \citenamefont {Liu}, \citenamefont {Swager},\ and\ \citenamefont
  {Dincă}}]{Campbell2015}%
  \BibitemOpen
  \bibfield  {author} {\bibinfo {author} {\bibfnamefont {M.~G.}\ \bibnamefont
  {Campbell}}, \bibinfo {author} {\bibfnamefont {S.~F.}\ \bibnamefont {Liu}},
  \bibinfo {author} {\bibfnamefont {T.~M.}\ \bibnamefont {Swager}}, \ and\
  \bibinfo {author} {\bibfnamefont {M.}~\bibnamefont {Dincă}},\ }\href
  {\doibase 10.1021/jacs.5b09600} {\bibfield  {journal} {\bibinfo  {journal}
  {J. Am. Chem. Soc.}\ }\textbf {\bibinfo {volume} {137}},\ \bibinfo {pages}
  {13780} (\bibinfo {year} {2015})}\BibitemShut {NoStop}%
\bibitem [{\citenamefont {Takenaka}\ \emph {et~al.}(2021)\citenamefont
  {Takenaka}, \citenamefont {Ishihara}, \citenamefont {Roppongi}, \citenamefont
  {Miao}, \citenamefont {Mizukami}, \citenamefont {Makita}, \citenamefont
  {Tsurumi}, \citenamefont {Watanabe}, \citenamefont {Takeya}, \citenamefont
  {Yamashita}, \citenamefont {Torizuka}, \citenamefont {Uwatoko}, \citenamefont
  {Sasaki}, \citenamefont {Huang}, \citenamefont {Xu}, \citenamefont {D.~Zhu},
  \citenamefont {Cheng}, \citenamefont {Shibauchi},\ and\ \citenamefont
  {Hashimoto}}]{Takenaka2021}%
  \BibitemOpen
  \bibfield  {author} {\bibinfo {author} {\bibfnamefont {T.}~\bibnamefont
  {Takenaka}}, \bibinfo {author} {\bibfnamefont {K.}~\bibnamefont {Ishihara}},
  \bibinfo {author} {\bibfnamefont {M.}~\bibnamefont {Roppongi}}, \bibinfo
  {author} {\bibfnamefont {Y.}~\bibnamefont {Miao}}, \bibinfo {author}
  {\bibfnamefont {Y.}~\bibnamefont {Mizukami}}, \bibinfo {author}
  {\bibfnamefont {T.}~\bibnamefont {Makita}}, \bibinfo {author} {\bibfnamefont
  {J.}~\bibnamefont {Tsurumi}}, \bibinfo {author} {\bibfnamefont
  {S.}~\bibnamefont {Watanabe}}, \bibinfo {author} {\bibfnamefont
  {J.}~\bibnamefont {Takeya}}, \bibinfo {author} {\bibfnamefont
  {M.}~\bibnamefont {Yamashita}}, \bibinfo {author} {\bibfnamefont
  {K.}~\bibnamefont {Torizuka}}, \bibinfo {author} {\bibfnamefont
  {Y.}~\bibnamefont {Uwatoko}}, \bibinfo {author} {\bibfnamefont
  {T.}~\bibnamefont {Sasaki}}, \bibinfo {author} {\bibfnamefont
  {X.}~\bibnamefont {Huang}}, \bibinfo {author} {\bibfnamefont
  {W.}~\bibnamefont {Xu}}, \bibinfo {author} {\bibfnamefont {N.~S.}\
  \bibnamefont {D.~Zhu}}, \bibinfo {author} {\bibfnamefont {J.-G.}\
  \bibnamefont {Cheng}}, \bibinfo {author} {\bibfnamefont {T.}~\bibnamefont
  {Shibauchi}}, \ and\ \bibinfo {author} {\bibfnamefont {K.}~\bibnamefont
  {Hashimoto}},\ }\href {\doibase DOI: 10.1126/sciadv.abf3996} {\bibfield
  {journal} {\bibinfo  {journal} {Sci. Adv.}\ }\textbf {\bibinfo {volume}
  {7}},\ \bibinfo {pages} {eabf3996} (\bibinfo {year} {2021})}\BibitemShut
  {NoStop}%
\bibitem [{\citenamefont {Wu}\ \emph {et~al.}(2017)\citenamefont {Wu},
  \citenamefont {Wang}, \citenamefont {Liu}, \citenamefont {Fu}, \citenamefont
  {Sun}, \citenamefont {Liu}, \citenamefont {Pan}, \citenamefont {Weng},
  \citenamefont {Dinca}, \citenamefont {Fu},\ and\ \citenamefont
  {Li}}]{Wu2014}%
  \BibitemOpen
  \bibfield  {author} {\bibinfo {author} {\bibfnamefont {M.}~\bibnamefont
  {Wu}}, \bibinfo {author} {\bibfnamefont {Z.}~\bibnamefont {Wang}}, \bibinfo
  {author} {\bibfnamefont {J.}~\bibnamefont {Liu}}, \bibinfo {author}
  {\bibfnamefont {W.~L.~H.}\ \bibnamefont {Fu}}, \bibinfo {author}
  {\bibfnamefont {L.}~\bibnamefont {Sun}}, \bibinfo {author} {\bibfnamefont
  {X.}~\bibnamefont {Liu}}, \bibinfo {author} {\bibfnamefont {M.}~\bibnamefont
  {Pan}}, \bibinfo {author} {\bibfnamefont {H.}~\bibnamefont {Weng}}, \bibinfo
  {author} {\bibfnamefont {M.}~\bibnamefont {Dinca}}, \bibinfo {author}
  {\bibfnamefont {L.}~\bibnamefont {Fu}}, \ and\ \bibinfo {author}
  {\bibfnamefont {J.}~\bibnamefont {Li}},\ }\href@noop {} {\bibfield  {journal}
  {\bibinfo  {journal} {2D Mater.}\ }\textbf {\bibinfo {volume} {4}},\ \bibinfo
  {pages} {015015} (\bibinfo {year} {2017})}\BibitemShut {NoStop}%
\bibitem [{\citenamefont {Hu}\ \emph {et~al.}(2014)\citenamefont {Hu},
  \citenamefont {Wang},\ and\ \citenamefont {Liu}}]{Hu2014}%
  \BibitemOpen
  \bibfield  {author} {\bibinfo {author} {\bibfnamefont {H.}~\bibnamefont
  {Hu}}, \bibinfo {author} {\bibfnamefont {Z.}~\bibnamefont {Wang}}, \ and\
  \bibinfo {author} {\bibfnamefont {F.}~\bibnamefont {Liu}},\ }\href@noop {}
  {\bibfield  {journal} {\bibinfo  {journal} {Nanoscale Res. Lett.}\ }\textbf
  {\bibinfo {volume} {9}},\ \bibinfo {pages} {1} (\bibinfo {year}
  {2014})}\BibitemShut {NoStop}%
\bibitem [{\citenamefont {Lima}\ \emph {et~al.}(2019)\citenamefont {Lima},
  \citenamefont {Ferreira},\ and\ \citenamefont {Miwa}}]{felipe45}%
  \BibitemOpen
  \bibfield  {author} {\bibinfo {author} {\bibfnamefont {F.~C.}\ \bibnamefont
  {Lima}}, \bibinfo {author} {\bibfnamefont {G.~J.}\ \bibnamefont {Ferreira}},
  \ and\ \bibinfo {author} {\bibfnamefont {R.~H.}\ \bibnamefont {Miwa}},\
  }\href@noop {} {\bibfield  {journal} {\bibinfo  {journal} {Phys. Chem. Chem.
  Phys.}\ }\textbf {\bibinfo {volume} {21}},\ \bibinfo {pages} {22344}
  (\bibinfo {year} {2019})}\BibitemShut {NoStop}%
\bibitem [{\citenamefont {Sun}\ \emph {et~al.}(2018)\citenamefont {Sun},
  \citenamefont {Tan}, \citenamefont {Feng}, \citenamefont {Zhao},\ and\
  \citenamefont {Petek}}]{Sun2018}%
  \BibitemOpen
  \bibfield  {author} {\bibinfo {author} {\bibfnamefont {H.}~\bibnamefont
  {Sun}}, \bibinfo {author} {\bibfnamefont {S.}~\bibnamefont {Tan}}, \bibinfo
  {author} {\bibfnamefont {M.}~\bibnamefont {Feng}}, \bibinfo {author}
  {\bibfnamefont {J.}~\bibnamefont {Zhao}}, \ and\ \bibinfo {author}
  {\bibfnamefont {H.}~\bibnamefont {Petek}},\ }\href {\doibase
  10.1021/acs.jpcc.8b03353} {\bibfield  {journal} {\bibinfo  {journal} {The
  Jour. of Phys. Chem. C}\ }\textbf {\bibinfo {volume} {122}},\ \bibinfo
  {pages} {18659} (\bibinfo {year} {2018})}\BibitemShut {NoStop}%
\bibitem [{\citenamefont {Wang}\ \emph {et~al.}(2013)\citenamefont {Wang},
  \citenamefont {Su},\ and\ \citenamefont {Liu}}]{Wang2013}%
  \BibitemOpen
  \bibfield  {author} {\bibinfo {author} {\bibfnamefont {Z.~F.}\ \bibnamefont
  {Wang}}, \bibinfo {author} {\bibfnamefont {N.}~\bibnamefont {Su}}, \ and\
  \bibinfo {author} {\bibfnamefont {F.}~\bibnamefont {Liu}},\ }\href {\doibase
  https://doi.org/10.1021/nl401147u} {\bibfield  {journal} {\bibinfo  {journal}
  {Nanoletters}\ }\textbf {\bibinfo {volume} {13}},\ \bibinfo {pages} {2842}
  (\bibinfo {year} {2013})}\BibitemShut {NoStop}%
\bibitem [{\citenamefont {Li}\ \emph {et~al.}(2020)\citenamefont {Li},
  \citenamefont {Zhirihin}, \citenamefont {Ni}, \citenamefont {Slobozhanyuk},
  \citenamefont {Alù},\ and\ \citenamefont {Khanikaev}}]{Khanikaev2020}%
  \BibitemOpen
  \bibfield  {author} {\bibinfo {author} {\bibfnamefont {M.}~\bibnamefont
  {Li}}, \bibinfo {author} {\bibfnamefont {D.}~\bibnamefont {Zhirihin}},
  \bibinfo {author} {\bibfnamefont {M.~G.~X.}\ \bibnamefont {Ni}}, \bibinfo
  {author} {\bibfnamefont {D.~F.~A.}\ \bibnamefont {Slobozhanyuk}}, \bibinfo
  {author} {\bibfnamefont {A.}~\bibnamefont {Alù}}, \ and\ \bibinfo {author}
  {\bibfnamefont {A.~B.}\ \bibnamefont {Khanikaev}},\ }\href {\doibase
  https://doi.org/10.1038/s41566-019-0561-9} {\bibfield  {journal} {\bibinfo
  {journal} {Nature Photonics}\ }\textbf {\bibinfo {volume} {14}},\ \bibinfo
  {pages} {89} (\bibinfo {year} {2020})}\BibitemShut {NoStop}%
\bibitem [{\citenamefont {Kirsch}\ \emph {et~al.}(2021)\citenamefont {Kirsch},
  \citenamefont {Zhang}, \citenamefont {Kremer}, \citenamefont {Maczewsky},
  \citenamefont {Ivanov}, \citenamefont {Kartashov}, \citenamefont {Torner},
  \citenamefont {Bauer},\ and\ \citenamefont {Szameit}}]{Kirsch2021}%
  \BibitemOpen
  \bibfield  {author} {\bibinfo {author} {\bibfnamefont {M.~S.}\ \bibnamefont
  {Kirsch}}, \bibinfo {author} {\bibfnamefont {Y.}~\bibnamefont {Zhang}},
  \bibinfo {author} {\bibfnamefont {M.}~\bibnamefont {Kremer}}, \bibinfo
  {author} {\bibfnamefont {L.~J.}\ \bibnamefont {Maczewsky}}, \bibinfo {author}
  {\bibfnamefont {S.~K.}\ \bibnamefont {Ivanov}}, \bibinfo {author}
  {\bibfnamefont {Y.~V.}\ \bibnamefont {Kartashov}}, \bibinfo {author}
  {\bibfnamefont {L.}~\bibnamefont {Torner}}, \bibinfo {author} {\bibfnamefont
  {D.}~\bibnamefont {Bauer}}, \ and\ \bibinfo {author} {\bibfnamefont
  {M.}~\bibnamefont {Szameit}, \bibfnamefont {A.~Heinrich}},\ }\href {\doibase
  https://doi.org/10.1038/s41567-021-01275-3} {\bibfield  {journal} {\bibinfo
  {journal} {Nature Physics}\ }\textbf {\bibinfo {volume} {17}},\ \bibinfo
  {pages} {995} (\bibinfo {year} {2021})}\BibitemShut {NoStop}%
\bibitem [{\citenamefont {Cho}\ \emph {et~al.}(2015)\citenamefont {Cho},
  \citenamefont {Deng}, \citenamefont {Miyasaka}, \citenamefont {Dong},
  \citenamefont {Cho}, \citenamefont {Neimark}, \citenamefont {Kang},
  \citenamefont {Yaghi},\ and\ \citenamefont {Terasaki}}]{Cho2015}%
  \BibitemOpen
  \bibfield  {author} {\bibinfo {author} {\bibfnamefont {H.~S.}\ \bibnamefont
  {Cho}}, \bibinfo {author} {\bibfnamefont {H.}~\bibnamefont {Deng}}, \bibinfo
  {author} {\bibfnamefont {K.}~\bibnamefont {Miyasaka}}, \bibinfo {author}
  {\bibfnamefont {Z.}~\bibnamefont {Dong}}, \bibinfo {author} {\bibfnamefont
  {M.}~\bibnamefont {Cho}}, \bibinfo {author} {\bibfnamefont {A.~V.}\
  \bibnamefont {Neimark}}, \bibinfo {author} {\bibfnamefont {J.~K.}\
  \bibnamefont {Kang}}, \bibinfo {author} {\bibfnamefont {O.~M.}\ \bibnamefont
  {Yaghi}}, \ and\ \bibinfo {author} {\bibfnamefont {O.}~\bibnamefont
  {Terasaki}},\ }\href@noop {} {\bibfield  {journal} {\bibinfo  {journal}
  {Nature}\ }\textbf {\bibinfo {volume} {527}},\ \bibinfo {pages} {503}
  (\bibinfo {year} {2015})}\BibitemShut {NoStop}%
\bibitem [{\citenamefont {Xu}\ \emph {et~al.}(2015)\citenamefont {Xu},
  \citenamefont {Gao},\ and\ \citenamefont {Jiang}}]{Xu2015}%
  \BibitemOpen
  \bibfield  {author} {\bibinfo {author} {\bibfnamefont {H.}~\bibnamefont
  {Xu}}, \bibinfo {author} {\bibfnamefont {J.}~\bibnamefont {Gao}}, \ and\
  \bibinfo {author} {\bibfnamefont {D.}~\bibnamefont {Jiang}},\ }\href@noop {}
  {\bibfield  {journal} {\bibinfo  {journal} {Nat. Chem.}\ }\textbf {\bibinfo
  {volume} {7}},\ \bibinfo {pages} {905} (\bibinfo {year} {2015})}\BibitemShut
  {NoStop}%
\bibitem [{\citenamefont {Dienstmaier}\ \emph {et~al.}(2011)\citenamefont
  {Dienstmaier}, \citenamefont {Gigler}, \citenamefont {Goetz}, \citenamefont
  {Knochel}, \citenamefont {T.~Bein}, \citenamefont {Reichlmaier},
  \citenamefont {Heckl},\ and\ \citenamefont {Lackinger}}]{Dienstmaier2011}%
  \BibitemOpen
  \bibfield  {author} {\bibinfo {author} {\bibfnamefont {J.~F.}\ \bibnamefont
  {Dienstmaier}}, \bibinfo {author} {\bibfnamefont {A.~M.}\ \bibnamefont
  {Gigler}}, \bibinfo {author} {\bibfnamefont {A.~J.}\ \bibnamefont {Goetz}},
  \bibinfo {author} {\bibfnamefont {P.}~\bibnamefont {Knochel}}, \bibinfo
  {author} {\bibfnamefont {A.~L.}\ \bibnamefont {T.~Bein}}, \bibinfo {author}
  {\bibfnamefont {S.}~\bibnamefont {Reichlmaier}}, \bibinfo {author}
  {\bibfnamefont {W.~M.}\ \bibnamefont {Heckl}}, \ and\ \bibinfo {author}
  {\bibfnamefont {M.}~\bibnamefont {Lackinger}},\ }\href@noop {} {\bibfield
  {journal} {\bibinfo  {journal} {ACS Nano}\ }\textbf {\bibinfo {volume} {5}},\
  \bibinfo {pages} {9737} (\bibinfo {year} {2011})}\BibitemShut {NoStop}%
\bibitem [{\citenamefont {Zhang}\ \emph {et~al.}(2013)\citenamefont {Zhang},
  \citenamefont {Tian}, \citenamefont {Hanifi}, \citenamefont {Zhang},
  \citenamefont {Sue}, \citenamefont {Zhou}, \citenamefont {Zhang},
  \citenamefont {Zhao}, \citenamefont {Liu},\ and\ \citenamefont
  {Li}}]{Zhang2013}%
  \BibitemOpen
  \bibfield  {author} {\bibinfo {author} {\bibfnamefont {K.-D.}\ \bibnamefont
  {Zhang}}, \bibinfo {author} {\bibfnamefont {J.}~\bibnamefont {Tian}},
  \bibinfo {author} {\bibfnamefont {D.}~\bibnamefont {Hanifi}}, \bibinfo
  {author} {\bibfnamefont {Y.}~\bibnamefont {Zhang}}, \bibinfo {author}
  {\bibfnamefont {A.~C.-H.}\ \bibnamefont {Sue}}, \bibinfo {author}
  {\bibfnamefont {T.-Y.}\ \bibnamefont {Zhou}}, \bibinfo {author}
  {\bibfnamefont {L.}~\bibnamefont {Zhang}}, \bibinfo {author} {\bibfnamefont
  {X.}~\bibnamefont {Zhao}}, \bibinfo {author} {\bibfnamefont {Y.}~\bibnamefont
  {Liu}}, \ and\ \bibinfo {author} {\bibfnamefont {Z.-T.}\ \bibnamefont {Li}},\
  }\href@noop {} {\bibfield  {journal} {\bibinfo  {journal} {J. Am. Chem.
  Soc.}\ }\textbf {\bibinfo {volume} {135}},\ \bibinfo {pages} {17913}
  (\bibinfo {year} {2013})}\BibitemShut {NoStop}%
\bibitem [{\citenamefont {Feldblyum}\ \emph {et~al.}(2015)\citenamefont
  {Feldblyum}, \citenamefont {McCreery}, \citenamefont {Andrews}, \citenamefont
  {Kurosawa}, \citenamefont {Santos}, \citenamefont {Duong}, \citenamefont
  {Fang}, \citenamefont {Ayzner},\ and\ \citenamefont {Bao}}]{Feldblyum2015}%
  \BibitemOpen
  \bibfield  {author} {\bibinfo {author} {\bibfnamefont {J.~I.}\ \bibnamefont
  {Feldblyum}}, \bibinfo {author} {\bibfnamefont {C.~H.}\ \bibnamefont
  {McCreery}}, \bibinfo {author} {\bibfnamefont {S.~C.}\ \bibnamefont
  {Andrews}}, \bibinfo {author} {\bibfnamefont {T.}~\bibnamefont {Kurosawa}},
  \bibinfo {author} {\bibfnamefont {E.~J.~G.}\ \bibnamefont {Santos}}, \bibinfo
  {author} {\bibfnamefont {V.}~\bibnamefont {Duong}}, \bibinfo {author}
  {\bibfnamefont {L.}~\bibnamefont {Fang}}, \bibinfo {author} {\bibfnamefont
  {A.~L.}\ \bibnamefont {Ayzner}}, \ and\ \bibinfo {author} {\bibfnamefont
  {Z.}~\bibnamefont {Bao}},\ }\href@noop {} {\bibfield  {journal} {\bibinfo
  {journal} {Chem. Commun.}\ }\textbf {\bibinfo {volume} {51}},\ \bibinfo
  {pages} {13894} (\bibinfo {year} {2015})}\BibitemShut {NoStop}%
\bibitem [{\citenamefont {Stock}\ and\ \citenamefont
  {Biswas}(2012)}]{Stock2012}%
  \BibitemOpen
  \bibfield  {author} {\bibinfo {author} {\bibfnamefont {N.}~\bibnamefont
  {Stock}}\ and\ \bibinfo {author} {\bibfnamefont {S.}~\bibnamefont {Biswas}},\
  }\href@noop {} {\bibfield  {journal} {\bibinfo  {journal} {Chem. Rev.}\
  }\textbf {\bibinfo {volume} {112}},\ \bibinfo {pages} {933} (\bibinfo {year}
  {2012})}\BibitemShut {NoStop}%
\bibitem [{\citenamefont {Tan}\ and\ \citenamefont {Zeng}(2019)}]{Tan2019}%
  \BibitemOpen
  \bibfield  {author} {\bibinfo {author} {\bibfnamefont {Y.~C.}\ \bibnamefont
  {Tan}}\ and\ \bibinfo {author} {\bibfnamefont {H.~C.}\ \bibnamefont {Zeng}},\
  }\href@noop {} {\bibfield  {journal} {\bibinfo  {journal} {ChemCatChem.}\
  }\textbf {\bibinfo {volume} {11}},\ \bibinfo {pages} {3138} (\bibinfo {year}
  {2019})}\BibitemShut {NoStop}%
\bibitem [{\citenamefont {Iqbal}\ \emph {et~al.}(2016)\citenamefont {Iqbal},
  \citenamefont {Abdelkareem}, \citenamefont {Sayed}, \citenamefont {Hamdan},
  \citenamefont {Baroutaji},\ and\ \citenamefont {Olabi}}]{Iqbal2016}%
  \BibitemOpen
  \bibfield  {author} {\bibinfo {author} {\bibfnamefont {A.}~\bibnamefont
  {Iqbal}}, \bibinfo {author} {\bibfnamefont {M.~A.}\ \bibnamefont
  {Abdelkareem}}, \bibinfo {author} {\bibfnamefont {E.~T.}\ \bibnamefont
  {Sayed}}, \bibinfo {author} {\bibfnamefont {N.~M.}\ \bibnamefont {Hamdan}},
  \bibinfo {author} {\bibfnamefont {A.}~\bibnamefont {Baroutaji}}, \ and\
  \bibinfo {author} {\bibfnamefont {A.}~\bibnamefont {Olabi}},\ }\href@noop {}
  {\bibfield  {journal} {\bibinfo  {journal} {Angewandte Chemie}\ }\textbf
  {\bibinfo {volume} {55}},\ \bibinfo {pages} {3566} (\bibinfo {year}
  {2016})}\BibitemShut {NoStop}%
\bibitem [{\citenamefont {Rizzo}\ \emph {et~al.}(2018)\citenamefont {Rizzo},
  \citenamefont {Veber}, \citenamefont {Cao}, \citenamefont {Bronner},
  \citenamefont {Chen}, \citenamefont {Zhao}, \citenamefont {Rodriguez},
  \citenamefont {Louie}, \citenamefont {Crommie},\ and\ \citenamefont
  {Fischer}}]{Rizzo2018}%
  \BibitemOpen
  \bibfield  {author} {\bibinfo {author} {\bibfnamefont {D.}~\bibnamefont
  {Rizzo}}, \bibinfo {author} {\bibfnamefont {G.}~\bibnamefont {Veber}},
  \bibinfo {author} {\bibfnamefont {T.}~\bibnamefont {Cao}}, \bibinfo {author}
  {\bibfnamefont {C.}~\bibnamefont {Bronner}}, \bibinfo {author} {\bibfnamefont
  {T.}~\bibnamefont {Chen}}, \bibinfo {author} {\bibfnamefont {F.}~\bibnamefont
  {Zhao}}, \bibinfo {author} {\bibfnamefont {H.}~\bibnamefont {Rodriguez}},
  \bibinfo {author} {\bibfnamefont {S.~G.}\ \bibnamefont {Louie}}, \bibinfo
  {author} {\bibfnamefont {M.~F.}\ \bibnamefont {Crommie}}, \ and\ \bibinfo
  {author} {\bibfnamefont {F.~R.}\ \bibnamefont {Fischer}},\ }\href {\doibase
  https://doi.org/10.1038/s41586-018-0376-8} {\bibfield  {journal} {\bibinfo
  {journal} {Nature}\ }\textbf {\bibinfo {volume} {560}},\ \bibinfo {pages}
  {204} (\bibinfo {year} {2018})}\BibitemShut {NoStop}%
\bibitem [{\citenamefont {Sheberla}\ \emph {et~al.}(2017)\citenamefont
  {Sheberla}, \citenamefont {Bachman}, \citenamefont {Elias}, \citenamefont
  {Sun}, \citenamefont {S.-Horn},\ and\ \citenamefont {Dinca}}]{Sheberla2017}%
  \BibitemOpen
  \bibfield  {author} {\bibinfo {author} {\bibfnamefont {D.}~\bibnamefont
  {Sheberla}}, \bibinfo {author} {\bibfnamefont {J.~C.}\ \bibnamefont
  {Bachman}}, \bibinfo {author} {\bibfnamefont {J.~S.}\ \bibnamefont {Elias}},
  \bibinfo {author} {\bibfnamefont {C.-J.}\ \bibnamefont {Sun}}, \bibinfo
  {author} {\bibfnamefont {Y.}~\bibnamefont {S.-Horn}}, \ and\ \bibinfo
  {author} {\bibfnamefont {M.}~\bibnamefont {Dinca}},\ }\href@noop {}
  {\bibfield  {journal} {\bibinfo  {journal} {Nature Materials}\ }\textbf
  {\bibinfo {volume} {16}},\ \bibinfo {pages} {220} (\bibinfo {year}
  {2017})}\BibitemShut {NoStop}%
\bibitem [{\citenamefont {Fang}\ \emph {et~al.}(2018)\citenamefont {Fang},
  \citenamefont {Zong},\ and\ \citenamefont {Mao}}]{Xang2018}%
  \BibitemOpen
  \bibfield  {author} {\bibinfo {author} {\bibfnamefont {X.}~\bibnamefont
  {Fang}}, \bibinfo {author} {\bibfnamefont {B.}~\bibnamefont {Zong}}, \ and\
  \bibinfo {author} {\bibfnamefont {S.}~\bibnamefont {Mao}},\ }\href {\doibase
  https://link.springer.com/article/10.1007/s40820-018-0218-0} {\bibfield
  {journal} {\bibinfo  {journal} {Nano-Micro Lett.}\ }\textbf {\bibinfo
  {volume} {10}},\ \bibinfo {pages} {112} (\bibinfo {year} {2018})}\BibitemShut
  {NoStop}%
\bibitem [{\citenamefont {Lim}\ \emph {et~al.}(2020)\citenamefont {Lim},
  \citenamefont {Fuchs}, \citenamefont {Piéchon},\ and\ \citenamefont
  {Montambaux}}]{Montambaux2020}%
  \BibitemOpen
  \bibfield  {author} {\bibinfo {author} {\bibfnamefont {L.-K.}\ \bibnamefont
  {Lim}}, \bibinfo {author} {\bibfnamefont {J.-N.}\ \bibnamefont {Fuchs}},
  \bibinfo {author} {\bibfnamefont {F.}~\bibnamefont {Piéchon}}, \ and\
  \bibinfo {author} {\bibfnamefont {G.}~\bibnamefont {Montambaux}},\ }\href
  {\doibase 10.1103/PhysRevB.101.045131} {\bibfield  {journal} {\bibinfo
  {journal} {Phys. Rev. B}\ }\textbf {\bibinfo {volume} {101}} (\bibinfo {year}
  {2020}),\ 10.1103/PhysRevB.101.045131}\BibitemShut {NoStop}%
\bibitem [{\citenamefont {Essafi}\ \emph {et~al.}(2017)\citenamefont {Essafi},
  \citenamefont {Jaubert},\ and\ \citenamefont {Udagawa}}]{Essafi2017}%
  \BibitemOpen
  \bibfield  {author} {\bibinfo {author} {\bibfnamefont {K.}~\bibnamefont
  {Essafi}}, \bibinfo {author} {\bibfnamefont {L.}~\bibnamefont {Jaubert}}, \
  and\ \bibinfo {author} {\bibfnamefont {M.}~\bibnamefont {Udagawa}},\
  }\href@noop {} {\bibfield  {journal} {\bibinfo  {journal} {Journal of
  Physics: Condensed Matter}\ }\textbf {\bibinfo {volume} {29}},\ \bibinfo
  {pages} {315802} (\bibinfo {year} {2017})}\BibitemShut {NoStop}%
\bibitem [{\citenamefont {Rhim}\ and\ \citenamefont {Yang}(2019)}]{Rhim2019}%
  \BibitemOpen
  \bibfield  {author} {\bibinfo {author} {\bibfnamefont {J.-W.}\ \bibnamefont
  {Rhim}}\ and\ \bibinfo {author} {\bibfnamefont {B.-J.}\ \bibnamefont
  {Yang}},\ }\href {\doibase 10.1103/PhysRevB.99.045107} {\bibfield  {journal}
  {\bibinfo  {journal} {Phys. Rev. B}\ }\textbf {\bibinfo {volume} {99}},\
  \bibinfo {pages} {045107} (\bibinfo {year} {2019})}\BibitemShut {NoStop}%
\bibitem [{\citenamefont {Liu}\ \emph {et~al.}(2009)\citenamefont {Liu},
  \citenamefont {Zhang}, \citenamefont {Wang},\ and\ \citenamefont
  {Li}}]{Liu2009}%
  \BibitemOpen
  \bibfield  {author} {\bibinfo {author} {\bibfnamefont {G.}~\bibnamefont
  {Liu}}, \bibinfo {author} {\bibfnamefont {P.}~\bibnamefont {Zhang}}, \bibinfo
  {author} {\bibfnamefont {Z.}~\bibnamefont {Wang}}, \ and\ \bibinfo {author}
  {\bibfnamefont {S.-S.}\ \bibnamefont {Li}},\ }\href {\doibase
  10.1103/PhysRevB.79.035323} {\bibfield  {journal} {\bibinfo  {journal} {Phys.
  Rev. B}\ }\textbf {\bibinfo {volume} {79}},\ \bibinfo {pages} {035323}
  (\bibinfo {year} {2009})}\BibitemShut {NoStop}%
\bibitem [{\citenamefont {Leykam}\ \emph {et~al.}(2018)\citenamefont {Leykam},
  \citenamefont {Andreanov},\ and\ \citenamefont {Flach}}]{Leykam2018}%
  \BibitemOpen
  \bibfield  {author} {\bibinfo {author} {\bibfnamefont {D.}~\bibnamefont
  {Leykam}}, \bibinfo {author} {\bibfnamefont {A.}~\bibnamefont {Andreanov}}, \
  and\ \bibinfo {author} {\bibfnamefont {S.}~\bibnamefont {Flach}},\ }\href
  {\doibase 10.1080/23746149.2018.1473052} {\bibfield  {journal} {\bibinfo
  {journal} {Advances in Physics: X}\ }\textbf {\bibinfo {volume} {3}},\
  \bibinfo {pages} {1473052} (\bibinfo {year} {2018})}\BibitemShut {NoStop}%
\bibitem [{\citenamefont {Bolens}\ and\ \citenamefont
  {Nagaosa}(2019)}]{Boleans2019}%
  \BibitemOpen
  \bibfield  {author} {\bibinfo {author} {\bibfnamefont {A.}~\bibnamefont
  {Bolens}}\ and\ \bibinfo {author} {\bibfnamefont {N.}~\bibnamefont
  {Nagaosa}},\ }\href {\doibase 10.1103/PhysRevB.99.165141} {\bibfield
  {journal} {\bibinfo  {journal} {Phys. Rev. B}\ }\textbf {\bibinfo {volume}
  {99}},\ \bibinfo {pages} {165141} (\bibinfo {year} {2019})}\BibitemShut
  {NoStop}%
\bibitem [{\citenamefont {Kim}\ \emph {et~al.}(2020)\citenamefont {Kim},
  \citenamefont {Mishra},\ and\ \citenamefont {Lee}}]{Kim2020}%
  \BibitemOpen
  \bibfield  {author} {\bibinfo {author} {\bibfnamefont {H.~S.}\ \bibnamefont
  {Kim}}, \bibinfo {author} {\bibfnamefont {A.}~\bibnamefont {Mishra}}, \ and\
  \bibinfo {author} {\bibfnamefont {S.}~\bibnamefont {Lee}},\ }\href {\doibase
  10.1103/PhysRevB.102.155113} {\bibfield  {journal} {\bibinfo  {journal}
  {Phys. Rev. B}\ }\textbf {\bibinfo {volume} {102}},\ \bibinfo {pages}
  {155113} (\bibinfo {year} {2020})}\BibitemShut {NoStop}%
\bibitem [{\citenamefont {Kang}\ \emph {et~al.}(2020)\citenamefont {Kang},
  \citenamefont {Fang}, \citenamefont {Ye}, \citenamefont {Po}, \citenamefont
  {Denlinger}, \citenamefont {Jozwiak}, \citenamefont {Bostwick}, \citenamefont
  {Rotenberg}, \citenamefont {Kaxiras}, \citenamefont {Checkelsky},\ and\
  \citenamefont {Comin}}]{Arpes2020}%
  \BibitemOpen
  \bibfield  {author} {\bibinfo {author} {\bibfnamefont {M.}~\bibnamefont
  {Kang}}, \bibinfo {author} {\bibfnamefont {S.}~\bibnamefont {Fang}}, \bibinfo
  {author} {\bibfnamefont {L.}~\bibnamefont {Ye}}, \bibinfo {author}
  {\bibfnamefont {H.~C.}\ \bibnamefont {Po}}, \bibinfo {author} {\bibfnamefont
  {J.}~\bibnamefont {Denlinger}}, \bibinfo {author} {\bibfnamefont
  {C.}~\bibnamefont {Jozwiak}}, \bibinfo {author} {\bibfnamefont
  {A.}~\bibnamefont {Bostwick}}, \bibinfo {author} {\bibfnamefont
  {E.}~\bibnamefont {Rotenberg}}, \bibinfo {author} {\bibfnamefont
  {E.}~\bibnamefont {Kaxiras}}, \bibinfo {author} {\bibfnamefont {J.~G.}\
  \bibnamefont {Checkelsky}}, \ and\ \bibinfo {author} {\bibfnamefont
  {R.}~\bibnamefont {Comin}},\ }\href {\doibase
  https://doi.org/10.1038/s41467-020-17465-1} {\bibfield  {journal} {\bibinfo
  {journal} {Nat. Commun.}\ }\textbf {\bibinfo {volume} {11}},\ \bibinfo
  {pages} {4004} (\bibinfo {year} {2020})}\BibitemShut {NoStop}%
\bibitem [{\citenamefont {Kane}\ and\ \citenamefont
  {Mele}(2005{\natexlab{a}})}]{Kane2005}%
  \BibitemOpen
  \bibfield  {author} {\bibinfo {author} {\bibfnamefont {C.~L.}\ \bibnamefont
  {Kane}}\ and\ \bibinfo {author} {\bibfnamefont {E.~J.}\ \bibnamefont
  {Mele}},\ }\href {\doibase 10.1103/PhysRevLett.95.226801} {\bibfield
  {journal} {\bibinfo  {journal} {Phys. Rev. Lett.}\ }\textbf {\bibinfo
  {volume} {95}},\ \bibinfo {pages} {226801} (\bibinfo {year}
  {2005}{\natexlab{a}})}\BibitemShut {NoStop}%
\bibitem [{\citenamefont {Kane}\ and\ \citenamefont
  {Mele}(2005{\natexlab{b}})}]{Kane_2005}%
  \BibitemOpen
  \bibfield  {author} {\bibinfo {author} {\bibfnamefont {C.~L.}\ \bibnamefont
  {Kane}}\ and\ \bibinfo {author} {\bibfnamefont {E.~J.}\ \bibnamefont
  {Mele}},\ }\href {\doibase 10.1103/PhysRevLett.95.146802} {\bibfield
  {journal} {\bibinfo  {journal} {Phys. Rev. Lett.}\ }\textbf {\bibinfo
  {volume} {95}},\ \bibinfo {pages} {146802} (\bibinfo {year}
  {2005}{\natexlab{b}})}\BibitemShut {NoStop}%
\bibitem [{\citenamefont {Rosales}\ \emph {et~al.}(2008)\citenamefont
  {Rosales}, \citenamefont {Pacheco}, \citenamefont {Barticevic}, \citenamefont
  {Latg{\'{e}}},\ and\ \citenamefont {Orellana}}]{Rosales2008}%
  \BibitemOpen
  \bibfield  {author} {\bibinfo {author} {\bibfnamefont {L.}~\bibnamefont
  {Rosales}}, \bibinfo {author} {\bibfnamefont {M.}~\bibnamefont {Pacheco}},
  \bibinfo {author} {\bibfnamefont {Z.}~\bibnamefont {Barticevic}}, \bibinfo
  {author} {\bibfnamefont {A.}~\bibnamefont {Latg{\'{e}}}}, \ and\ \bibinfo
  {author} {\bibfnamefont {P.~A.}\ \bibnamefont {Orellana}},\ }\href {\doibase
  10.1088/0957-4484/19/6/065402} {\bibfield  {journal} {\bibinfo  {journal}
  {Nanotechnology}\ }\textbf {\bibinfo {volume} {19}},\ \bibinfo {pages}
  {065402} (\bibinfo {year} {2008})}\BibitemShut {NoStop}%
\bibitem [{\citenamefont {Carrillo-Bastos}\ \emph {et~al.}(2016)\citenamefont
  {Carrillo-Bastos}, \citenamefont {Le\'on}, \citenamefont {Faria},
  \citenamefont {Latg\'e}, \citenamefont {Andrei},\ and\ \citenamefont
  {Sandler}}]{carrillo2016strained}%
  \BibitemOpen
  \bibfield  {author} {\bibinfo {author} {\bibfnamefont {R.}~\bibnamefont
  {Carrillo-Bastos}}, \bibinfo {author} {\bibfnamefont {C.}~\bibnamefont
  {Le\'on}}, \bibinfo {author} {\bibfnamefont {D.}~\bibnamefont {Faria}},
  \bibinfo {author} {\bibfnamefont {A.}~\bibnamefont {Latg\'e}}, \bibinfo
  {author} {\bibfnamefont {E.~Y.}\ \bibnamefont {Andrei}}, \ and\ \bibinfo
  {author} {\bibfnamefont {N.}~\bibnamefont {Sandler}},\ }\href {\doibase
  10.1103/PhysRevB.94.125422} {\bibfield  {journal} {\bibinfo  {journal} {Phys.
  Rev. B}\ }\textbf {\bibinfo {volume} {94}},\ \bibinfo {pages} {125422}
  (\bibinfo {year} {2016})}\BibitemShut {NoStop}%
\bibitem [{\citenamefont {Torres}\ \emph {et~al.}(2018)\citenamefont {Torres},
  \citenamefont {Faria},\ and\ \citenamefont {Latg\'e}}]{torres2018}%
  \BibitemOpen
  \bibfield  {author} {\bibinfo {author} {\bibfnamefont {V.}~\bibnamefont
  {Torres}}, \bibinfo {author} {\bibfnamefont {D.}~\bibnamefont {Faria}}, \
  and\ \bibinfo {author} {\bibfnamefont {A.}~\bibnamefont {Latg\'e}},\ }\href
  {\doibase 10.1103/PhysRevB.97.165429} {\bibfield  {journal} {\bibinfo
  {journal} {Phys. Rev. B}\ }\textbf {\bibinfo {volume} {97}},\ \bibinfo
  {pages} {165429} (\bibinfo {year} {2018})}\BibitemShut {NoStop}%
\bibitem [{\citenamefont {Leon}\ \emph {et~al.}(2019)\citenamefont {Leon},
  \citenamefont {Costa}, \citenamefont {Chico},\ and\ \citenamefont
  {Latgé}}]{Leon2019}%
  \BibitemOpen
  \bibfield  {author} {\bibinfo {author} {\bibfnamefont {C.}~\bibnamefont
  {Leon}}, \bibinfo {author} {\bibfnamefont {M.}~\bibnamefont {Costa}},
  \bibinfo {author} {\bibfnamefont {L.}~\bibnamefont {Chico}}, \ and\ \bibinfo
  {author} {\bibfnamefont {A.}~\bibnamefont {Latgé}},\ }\href@noop {}
  {\bibfield  {journal} {\bibinfo  {journal} {Scientific Reports}\ }\textbf
  {\bibinfo {volume} {9}},\ \bibinfo {pages} {1} (\bibinfo {year}
  {2019})}\BibitemShut {NoStop}%
\bibitem [{\citenamefont {Sheng}\ \emph {et~al.}(2006)\citenamefont {Sheng},
  \citenamefont {Weng}, \citenamefont {Sheng},\ and\ \citenamefont
  {Haldane}}]{Sheng2006}%
  \BibitemOpen
  \bibfield  {author} {\bibinfo {author} {\bibfnamefont {D.~N.}\ \bibnamefont
  {Sheng}}, \bibinfo {author} {\bibfnamefont {Z.~Y.}\ \bibnamefont {Weng}},
  \bibinfo {author} {\bibfnamefont {L.}~\bibnamefont {Sheng}}, \ and\ \bibinfo
  {author} {\bibfnamefont {F.~D.~M.}\ \bibnamefont {Haldane}},\ }\href
  {\doibase 10.1103/PhysRevLett.97.036808} {\bibfield  {journal} {\bibinfo
  {journal} {Phys. Rev. Lett.}\ }\textbf {\bibinfo {volume} {97}},\ \bibinfo
  {pages} {036808} (\bibinfo {year} {2006})}\BibitemShut {NoStop}%
\end{thebibliography}%

\end{document}